\documentclass[preprint]{aastex}
\slugcomment{Submitted to ApJL}
\shortauthors{Zheng et al.}
\begin{document}

\title{The initial morphologies of the wavefronts of extreme ultraviolet waves}
\author{Ruisheng Zheng$^{1}$, Zhike Xue$^{2}$, Yao Chen$^{1}$, Bing Wang$^{1}$, and Hongqiang Song$^{1}$}
\affil{$^{1}$Shandong Provincial Key Laboratory of Optical Astronomy and Solar-Terrestrial Environment, and Institute of Space Sciences, Shandong University, 264209 Weihai, China; ruishengzheng@sdu.edu.cn\\
 $^{2}$Yunnan Observatories, Chinese Academy of Sciences, Kunming 650216, China\\}

\begin{abstract}
The morphologies of the wavefronts of extreme ultraviolet (EUV) waves can shed light on their physical nature and driving mechanism that are still strongly debated. In reality, the wavefronts always deform after interacting with ambient coronal structures during their propagation. Here, we focus on the initial wavefront morphologies of four selected EUV waves that are closely associated with jets or flux rope eruptions, using the high spatio-temporal resolution observations and different perspectives from the Solar Dynamics Observatory and the Solar-Terrestrial Relations Observatory. For the jet-driven waves, the jets originated from one end of the overlying closed loops, and the arc-shaped wavefront formed around the other far end of the expanding loops. The extrapolated field lines of the Potential Field Source Surface model show the close relationships between the jets, the wavefronts, and the overlying closed loops. For the flux-rope-driven waves, the flux ropes (sigmoids) lifted off beneath the overlying loops, and the circular wavefronts had an intimate spatio-temporal relation with the expanding loops. All the results suggest that the configuration of the overlying loops and their locations relative to the erupting cores are very important for the formation and morphology of the wavefronts, and both two jet-driven waves and two flux-rope-driven waves are likely triggered by the sudden expansion of the overlying closed loops. We also propose that the wavefront of EUV wave is possibly integrated by a chain of wave components triggered by a series of separated expanding loops.

\end{abstract}

\keywords{Sun: activity --- Sun: corona --- Sun: oscillations --- waves}

\section{Introduction}
Extreme ultraviolet (EUV) waves are spectacular propagating disturbances in the solar corona. They are not only closely associated with various solar activities, such as coronal mass ejection (CMEs), solar flares, filament eruptions, coronal jets, type II radio bursts, and oscillations of ambient coronal structures, but also carry important information on the coronal energy transportation and the coronal magnetic fields, {\bf which is very useful for the global coronal seismology (Ballai 2007; Ballai et al. 2008; Kwon et al. 2013; Long et al. 2013, 2017).}

Since its discovery (Moses et al. 1997; Thompson et al. 1998), the EUV wave has been strongly debated on its physical nature and driving mechanism, and has been interpreted in different models (Thompson et al. 1998; Chen et al. 2002; Zhukov \& Auch\`{e}re 2004; Ballai et al. 2005; Attrill et al. 2007; Wang et al. 2009; Zheng et al. 2012). Using the high spatio-temporal resolution and full-disk, wide-temperature coverage observations from the Atmospheric Imaging Assembly (AIA; Lemen et al. 2012) on board the Solar Dynamics Observatory (SDO; Pesnell et al. 2012), hundreds of EUV waves have been studied in details (Nitta et al. 2013; Liu \& Ofman 2014). More and more observational and numerical evidences show that an EUV wave consists of a bimodal composition of an outer fast-mode magnetohydrodynamic wave and an inner non-wave CME component (Liu et al. 2010; Chen \& Wu 2011; Downs et al. 2012), and it is generally believed that the onset of the EUV wave strongly depends on the lateral expansion of the CME flanks (Patsourakos \& Vourlidas 2012; Cheng et al. 2012; Liu \& Ofman 2014; Warmuth 2015).

For typical CME-related EUV waves, they always have driving core of erupting flux ropes (filaments, sigmoids, coronal cavities and so on), and we here refer to them as flux-rope-driven waves. In fact, there is another kind of EUV waves that is primarily activated by coronal jets, and we call them as jet-driven waves. Zheng et al. (2012a) first reported a small-scale jet-driven EUV wave without any CME during a failed eruption, and proposed that any kind of mass ejection could possibly drive an EUV wave. Zheng et al. (2012b) also showed four homologous EUV waves associated closely with recurrent jets. Recently, more jet-driven waves have been reported in details (Shen et al. 2018a, 2018b).

The wavefronts of EUV waves are best seen as intensity enhancements in AIA 193 and 211~{\AA}, which have the peak temperature responses at 1.6-2.0 MK. In reality, the EUV waves travel continuously from the eruption center, and the wavefronts always encounter various magnetic structures, such as active regions (ARs), coronal holes, coronal cavities, coronal loops, and streamers. Therefore, EUV waves usually occurs reflections, refractions, transmissions, and the wave-mode conversion (Li et al. 2012; Olmedo et al. 2012; Shen et al. 2013; Xue et al. 2013; Chen et al. 2016; Zong \& Dai 2017; Zheng et al. 2018), and the wavefronts become irregular and fragmented. Except the interaction with the ambient complex coronal structures, the initial morphology of EUV waves can be classified as a full circle or an arc. What causes the different initial morphologies of wavefronts of different EUV waves?

On the other hand, the wavefronts of flux-rope-driven waves usually travel continuously and quasi-isotropically from the eruption center as a circle, and those of the jet-driven waves always have propagates in limited angular extent as a brow/arc. Do the different initial wavefront morphologies depend on the different eruptions? How the various wavefront morphologies form? Do the diverse wavefronts have the same trigger mechanism? To our knowledge, the initial morphologies of the wavefronts have not been fully noticed by previous studies. Here we first study the causes of the various initial morphologies of the wavefronts. Two jet-driven waves and two flux-rope-driven waves are selected to understand the physical nature and formation mechanism of diverse initial wavefronts.

\section{Observations}
All the selected waves were simultaneously captured by different perspectives of the SDO and twin (A and B) spacecrafts of Solar Terrestrial Relations Observatory (STEREO; Kaiser et al. 2008). We mainly use the EUV observations from the AIA on SDO and from the Extreme Ultraviolet Imager (EUVI; Howard et al. 2008) on the STEREO-A and STEREO-B. The AIA instrument's seven EUV wavelengths involve a wide range of temperatures. The AIA images cover the full disk (4096~$\times$4096 pixels) of the Sun and up to 0.5 $R_\odot$ above the limb, with a pixel resolution of 0.6$"$ and a cadence of 12 s. The EUVI images have a pixel resolution of 1.58$"$, and their cadences are 5 minutes or 10 minutes for the 171~{\AA}, 195~{\AA} and 304~{\AA}. The running-ratio-difference EUV images are used to display better the wavefronts and other erupting structures.

In addition, Magnetograms from the Helioseismic and Magnetic Imager (HMI; Scherrer et al. 2012)on SDO, with a cadence of 45 s and pixel scale of 0.6$"$, were used to check the magnetic field configuration of the source region. The kinematics of the jets and associated waves are analyzed with the time-slice approach. We also employ the Potential Field Source Surface (PFSS; Schrijver {\&} De Rosa 2003) model to extrapolate the coronal magnetic field. The coronal magnetic filed lines are extrapolated with some main codes, {\it pfss\_trace\_field.pro} and {\it pfss\_draw\_field.pro}, in the PFSS package of SolarSoftWare (see the template routine of {\it pfss\_sample1.pro}). The speeds and accelerations with errors in the context are obtained by fitting with linear ({\it linfit.pro}) and quadratic ({\it poly\_fit.pro}) functions, assuming a measurement uncertainty of 4 pixels ($\sim1.74$ Mm) for the selected points.

\section{Results}
\subsection{wavefronts of jet-driven waves}
The first jet-driven wave occurred in AR 11149 on 2011 January 27 (Figure 1 and its animation). It was a typical jet (the yellow arrow), and ambient coronal loops (the blue arrow) were very clear in AIA 304 and 171~{\AA}, respectively (panels (a)-(b)). The jet was captured in the limb view of AIA and the disk perspective of EUVI-A, and was much more apparent in running-ratio difference images of AIA 193~{\AA} and EUVI-A 195~{\AA} (panels (c)-(d)). As a result, a coronal wave formed ahead of the southward jet, and showed an arc-shaped wavefront (red arrows in panels (e)-(f)). Superposed with the extrapolated field lines by the PFSS model, it is clear that the jet emanated from the north footpoints of extrapolated lines, and the arc-shaped wavefront just formed around the south footpoints of extrapolated lines. During its propagation, the bright arc-shaped wavefront became diffuse and faint (red arrows in panels (g)-(h)).  In the time-slice plot (the panel (g1)) along the jet-wave trajectory in AIA 193~{\AA} (S1; the green dashed line in panel (g)), the wave front appeared as an inclined bright strip. By fitting the green and red asterisks with linear and quadratic functions, the results show that the speed of the jet was 280$\pm$8 km s$^{-1}$, and the wave decelerated from 462$\pm$13 to 203$\pm$13 km s$^{-1}$, with an acceleration of -0.298$\pm$0.019 km s$^{-2}$. {\bf The negative acceleration is comparable to the results (ranging from $\sim-0.29$ to $\sim-0.06$ km s$^{-2}$) in the previous statistical investigation (Long et al. 2011) for EUV waves with only STEREO observations.}

Another jet-driven wave happened in AR 12017 on 2014 March 28 (Figure 2 and its animation). The jet (the yellow arrow) emanated from the negative (blue contours) leading sunspot enclosing some parasitic positive (green contours) polarities (panel (a)). In the eruption center, the associated M2.0 flare had circular ribbons (white arrows), and some coronal loops (the blue arrow) extended out to remote positive magnetic polarities (panels (a)-(b)). The event is similar to the circular-ribbon flare with fan-spine-null-point configuration in Wang et al. (2012), and the extended loops are corresponding to the outer spine in this configuration that is confirmed by the PFSS extrapolated field lines (superposed blue lines in panel (e)). Following the northward jet, a coronal wave formed mainly in the north, and the arc-shaped wavefront appeared bright and dark in AIA 193 and 171~{\AA}, respectively (red arrows in panels (c)-(d)). The position of the wavefront is consistent with the north footpoints of extrapolated field lines, and the arc shape is likely due to the wide range of the outer footpoints of a series of spine loops. In the limb perspective of EUVI-A, the jet (the yellow arrow) clearly moved upward along the spine, and the wavefront (the red arrow) was just ahead of the jet (panel (f)). In the time-slice plot (panel (g1)) along the jet-wave trajectory in AIA 193~{\AA} (S2; the green dotted line in panel (g)), the wave front showed clearly as an inclined bright strip. Following the expanding, the east and west flanks of the arc-shaped wavefront in AIA are corresponding to the southern and northern parts in EUVI-A (red arrows in panel (h)). By fitting the green and red asterisks with linear and quadratic functions, the jet had a speed of 207$\pm$11 km s$^{-1}$, and the wave had a high initial speed of 579$\pm$79 and decreased quickly to 205$\pm$79 km s$^{-1}$, with an acceleration of -0.822$\pm$0.056 km s$^{-2}$. {\bf The negative acceleration is much higher than the results (ranging from $\sim-0.29$ to $\sim-0.06$ km s$^{-2}$) of Long et al. (2011), which is possibly due to the much higher spatio-temporal resolution of the SDO observations.}

\subsection{Wavefronts of flux-rope-driven waves}
The first flux-rope-driven wave was closely associated with a sigmoid eruption in the quiet Sun on 2018 February 24 (Figure 3 and its animation). The sigmoidal filament (the white arrow) and two J-shaped loops (blue arrows) were evident in AIA 304 and 193~{\AA}, respectively (panels (a)-(b)). Note that there was a cluster of coronal loops (L1; the black arrow) to the west of the eruption region (panel (b)). After the eruption, there formed a circular wave (red arrows) that appeared as an arc in the limb view of EUVI-A (panels (c)-(f)). Due to the hindrance of the L1 in the west, the northwestern part of the wavefront was a bit of faint (the red arrow in panel (g)). In the time-slice plots (panels (h)-(i)) in AIA 193 and 171~{\AA} along a selected wave trajectory (S3; the green dashed line in panel (g)), it is clear that the wavefront (red arrows) was intimately related to the expanding loops (blue arrows). By fitting the green and red asterisks with a linear function, the northeastward loop expansion speed is 85$\pm$12 km s$^{-1}$, and the wave had constant speeds of 209$\pm$8 and 218$\pm$11 km s$^{-1}$ in northeastward and southwestward directions, respectively.

The second flux-rope-driven wave happened in a backside AR on 2012 March 26, and was observed on the disk by EUVI-B and in the limb by AIA, respectively (Figure 4 and its animation). In the core of the AR, the sigmoidal filament (the white arrow) in EUVI-B 304~{\AA} indicating the existence of a flux rope (panel (a) and (a1)). Due to the limited resolution of EUVI-B, the J-shaped loops of sigmoid and the overlying loops were faint. But it is very clear for the twin deep dimmings (twin yellow arrows) and the loop-like dimmings (northward yellow arrows), as typical productions of the sigmoid eruption (panels (a)-(b)). The expanding overlying loops were clearly seen in running-ratio-difference images in 171~{\AA} of EUVI-B and AIA (panels (c)-(d)). In the limb perspective of AIA, the expanding loops were obviously stratified, and the outermost loops were likely corresponding to the upward expanding loops in disk view (blue arrows in panels (c)-(d)). On the other hand, the eruption was also accompanied with a circular EUV wave in EUVI-B that only appeared as a projected arc in AIA (red arrows in panels (e)-(h)). We chose an limb arc (S4) of the altitude of 1.15 $R_\odot$ in AIA to analyse the kinematics of the wavefront. In the time-slice plots in AIA 171 and 193~{\AA} (panels (d1) and (f1)), it is clear that the wavefront (green dotted lines) had a close spatio-temporal relationship with the expanding loops (red dotted lines). By fitting the green and red asterisks with a linear function, the loop expansion speeds are 112$\pm$10 and 175$\pm$6 km s$^{-1}$, and the wave had constant speeds of 234$\pm$6 and 226$\pm$5 km s$^{-1}$ in northward and southward directions, respectively.

\section{Conclusions and Discussion}
Using high quality observations and different perspectives from the SDO and STEREO, we study the initial wavefront morphologies of jet-driven and flux-rope-driven EUV waves. Without consideration of the interactions with ambient coronal structures for the four examples, the wavefronts of jet-driven waves are arc-shaped propagating in a limited angular extent (Figure 1 and 2), and the wavefronts of flux-rope-driven waves traveled quasi-isotropically from the eruption center in a full circle (Figure 3 and 4).

Figure 5 shows the running-ratio intensity profiles of wavefronts in in AIA 193~{\AA} or EUVI 195~{\AA} along some individual paths (green dotted lines: S1a, S2, S3 and S4a). The peaks (red pluses) and outermost edges (blue pluses) of wavefronts are distinguished from the around stationary brightenings by comparing with wave signals in running-ratio-difference images (red and blue pluses in Figures 1-4). Then, we derived the amplitudes and the frontal part widths of each wave, from the relative intensities of profile peaks (red pluses) and the distances from peaks to outermost edges (two vertical lines). {\bf It is obvious that the wave amplitudes are decreasing (red numbers), which is consistent with the wavefront evolution of fast-mode waves (Warmuth et al. 2001; Veronig et al. 2010; Warmuth \& Mann 2011). The frontal part widths (blue numbers) increased by $\sim$30-150 Mm in 10 Minutes, except for the increase of $\sim$30 Mm in 4 minutes for the small-scale wave on 2018 February 24, which is in accord with the statistical results of Long et al. (2011) that showed the increase of $\sim$160 Mm in 20 minutes for spatial widths of EUV waves (Figures 12-14 therein).} Note that because some amplitudes of waves are possibly affected by the another jet (the arrow in Figure 1(h)) and stationary loop brightenings (black pluses in Figure 3(f)-(g)), the correlation between the amplitude and speed of the short-duration waves is not clear. On the other hand, the jet-driven waves had large initial speeds (460 and 580 km s$^{-1}$) followed by decelerations, and the flux-rope-driven waves propagated nearly constant velocities (210-230 km s$^{-1}$). All the wave speeds are in the range of fast-mode waves (200-1500 km s$^{-1}$; Wills-Davey et al. 2007). {\bf Therefore, the characteristics of deceleration, pulse widening, and amplitude decrease suggest that the four waves in this letter are likely fast-mode dispersive magneto-acoustic waves.}

Though the speeds of the jets and expanding loops are almost two times smaller ($\sim$200 and $\sim$100 km s$^{-1}$) than their associated waves, the time-slice plots show clearly that the waves have a close spatio-temporal relationship with their associated jets and expanding loops. In addition, the extrapolated PFSS field lines connect the jet emanating from one footpoint and the wavefront forming at another far end. The above results show that the initial wavefront morphologies of EUV waves are closely associated with the overlying loops, and the waves are likely triggered by the sudden expansion of the loops that were abruptly concussed by the erupting cores (jets or flux ropes).

Obviously, the overlying loops act as a magnetic cage (Amari et al. 2018) to envelop the erupting core (jets or flux ropes). If there is no overlying loop, the jet should be released easily to the space, like the jet moving along open field lines in the coronal hole. Therefore, the overlying closed loops are likely necessary for the formation of the jet-driven waves. Furthermore, the jet usually originates from one compact end of the magnetic cage, and drives the magnetic cage to expand in the same direction; Thus, the wavefront mainly appears ahead of the magnetic cage in the particular angular extent. If the other end of the magnetic cage consists of many separate loop footpoints, the wavefront should be a larger arc, like the jet-driven wave in the fan-spine-null-point configuration (Figure 2). The flux rope explodes beneath the magnetic cage with two dispersed ends, and pushes the magnetic cage to expand outwards in all directions; As a result, the wavefront is circular. Accordingly, the topology of the overlying cage-like loops and their locations relative to the erupting core (jets or flux ropes) is potentially the key on the initial wavefront morphology of the EUV wave.

Based on the initial morphologies of the wavefronts of jet-driven and flux-rope-driven waves, we propose a scenario of the various wavefront formation of EUV waves (Figure 6). In the simple jet case (panel (a)), the jet (the pink patch) emanates from one compact end of the circumambient loops (the green lines) and pounds on some point of the overlying loops (two green dashed lines); Consequently, the arc-shaped wavefront (the blue diffuse shade) formed ahead of the expanding loops (top two green lines). For the jets in the fan-spine-null-point configuration (panel (b)), the jet (the pink patch) starts from the null point and hit some separate overlying loops (two dotted green lines), and therefore the separate expanding loops produce dispersed wavefronts (blue smooth shades). For the flux rope (the black circle and black S-shaped rope) eruption in the views of limb and disk (panels (c)-(d)), the eruption thrusts the overlying loops outwards vertically (green lines) and horizontally (red lines), and therefore the isolated wavefronts (blue smooth shades) are generated ahead of separate expanding loops. All the dispersed wavefronts constitute an entire diffuse wavefront (blue diffuse shades in panels (b) and (d)) that are usually observed. Hence, we suggest that the wavefront of EUV wave is possibly as a result of the integration of a chain of wave components triggered by a series of separated expanding loops.

{\bf In summary, the clear observational evidences and the magnetic field extrapolation results suggest that two jet-driven waves and two flux-rope-driven waves are likely fast-mode dispersive magneto-acoustic waves,} and they are all likely triggered by the sudden expansion of the overlying closed loops. The cage-like closed loops are important for the formation of EUV waves, and the configuration of the overlying loops and their locations relative to the erupting core are likely critical for the initial wavefront morphologies of EUV waves. We propose that the wavefront of EUV wave possibly consists of a chain of wave components triggered by a series of separate loops. Further more and better observations will be necessary to verify the results of the various wavefront morphologies of EUV waves in the present study.

\acknowledgments
SDO is a mission of NASA's Living With a Star Program. The authors thank the teams of SDO and STEREO for providing the data. This work is supported by grants NSFC 41331068, 11303101, and 11603013, Shandong Province Natural Science Foundation ZR2016AQ16, and Young Scholars Program of Shandong University, Weihai, 2016WHWLJH07. H. Q. Song is supported by the Natural Science Foundation of Shandong Province JQ201710.

\clearpage

\begin{figure}
\epsscale{0.6} \plotone{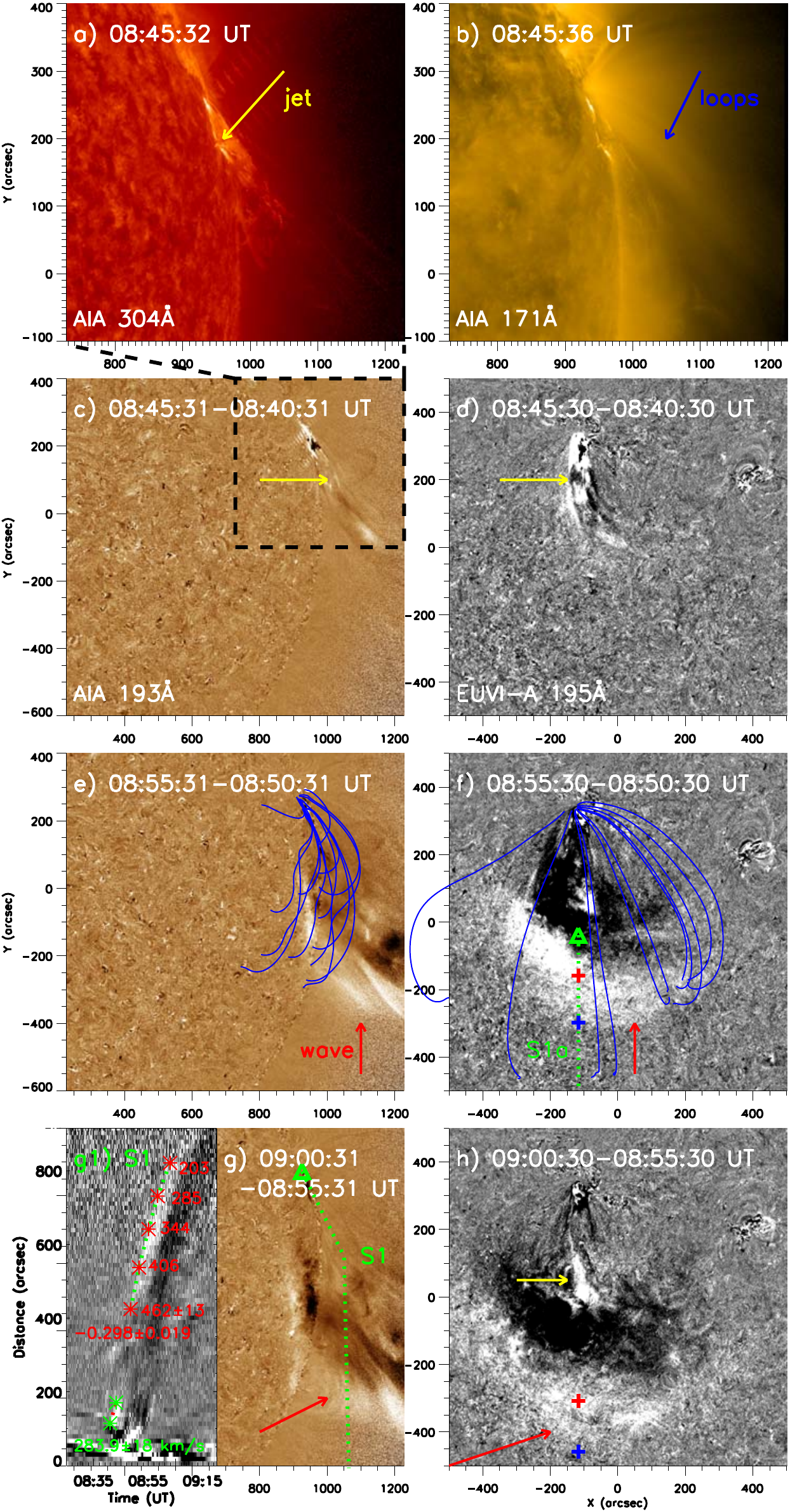}
\caption{The jet-driven wave on 2011 January 27. (a-b) The jet (the yellow arrow) and overlying loops (the blue arrow) in direct images of AIA 304 and 171~{\AA}. (c-h) The jet (yellow arrows) and associated EUV wave (red arrows) in running-ratio-difference images in AIA 193~{\AA} and EUVI-A 195~{\AA}. The dashed box in (c) indicates the field of view (FOV) of (a-b). The PFSS extrapolated field lines (blue) showing the overlying closed loops. (g1) The time-slice plot of running-ratio-difference AIA 193~{\AA} images showing the jet (yellow) and the EUV wave (red) along the green dotted line (S1 in (g)). The red and green asterisks connected by dotted lines are used to derive the wave and jet speed.
An animation of panels a-d is available. In all four panels the video begins at approximately 08:35:30 UT and ends at 09:05:10 UT. The video duration is 3 seconds.
\label{f1}}
\end{figure}

\clearpage

\begin{figure}
\epsscale{0.55}
\plotone{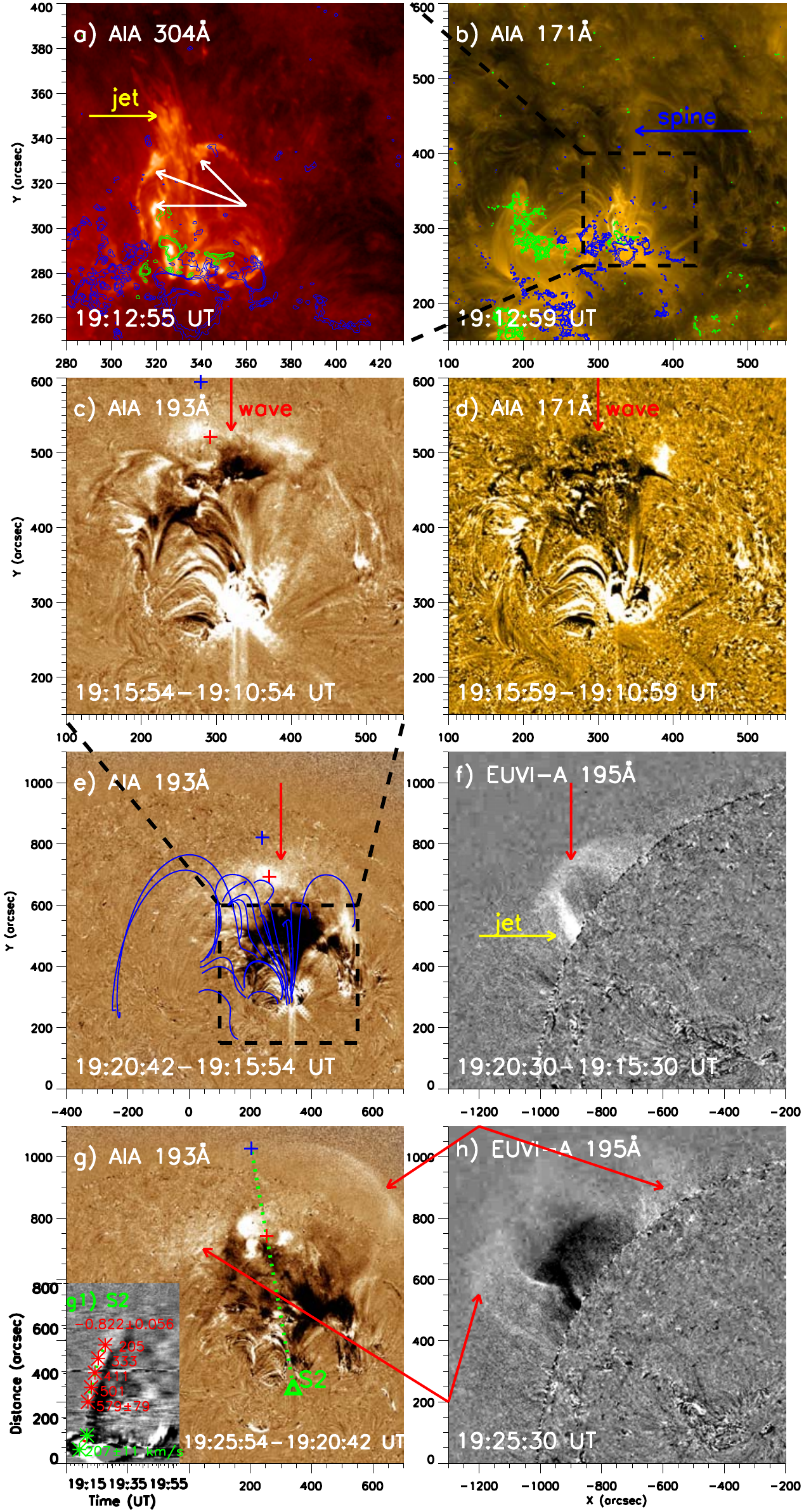}
\caption{The jet-driven wave on 2014 March 28. (a-b) The jet (the yellow arrow), the overlying spine loops (the blue arrow), and the circular flare ribbons (white arrows) in direct images of AIA 304 and 171~{\AA}. Contours of HMI line-of-sight magnetic fields are superposed with positive (negative) fields in green (blue), and the levels are 50, 100, and 150 G, respectively. The dashed box in (b) indicates the FOV of (a). (c-h) The jet (the yellow arrow) and associated EUV wave (red arrows) in running-ratio-difference images in AIA 193 and 171~{\AA} and EUVI-A 195~{\AA}. The dashed box in (e) indicates the FOV of (b-d). The PFSS extrapolated field lines (blue) in (e) showing the fan-spine-null-point configuration. (g1) The time-slice plot of running-ratio-difference AIA 193~{\AA} images showing the jet (yellow) and the EUV wave (red) along the green dotted line (S2 in (g)). The red and green asterisks connected by dotted lines are used to derive the wave and jet speed.
An animation of panels b, a, c, and f is available. In all four panels the video begins at approximately 19:13 UT and ends at 19:37 UT. The video duration is 2 seconds.
\label{f2}}
\end{figure}

\clearpage

\begin{figure}
\epsscale{0.6}
\plotone{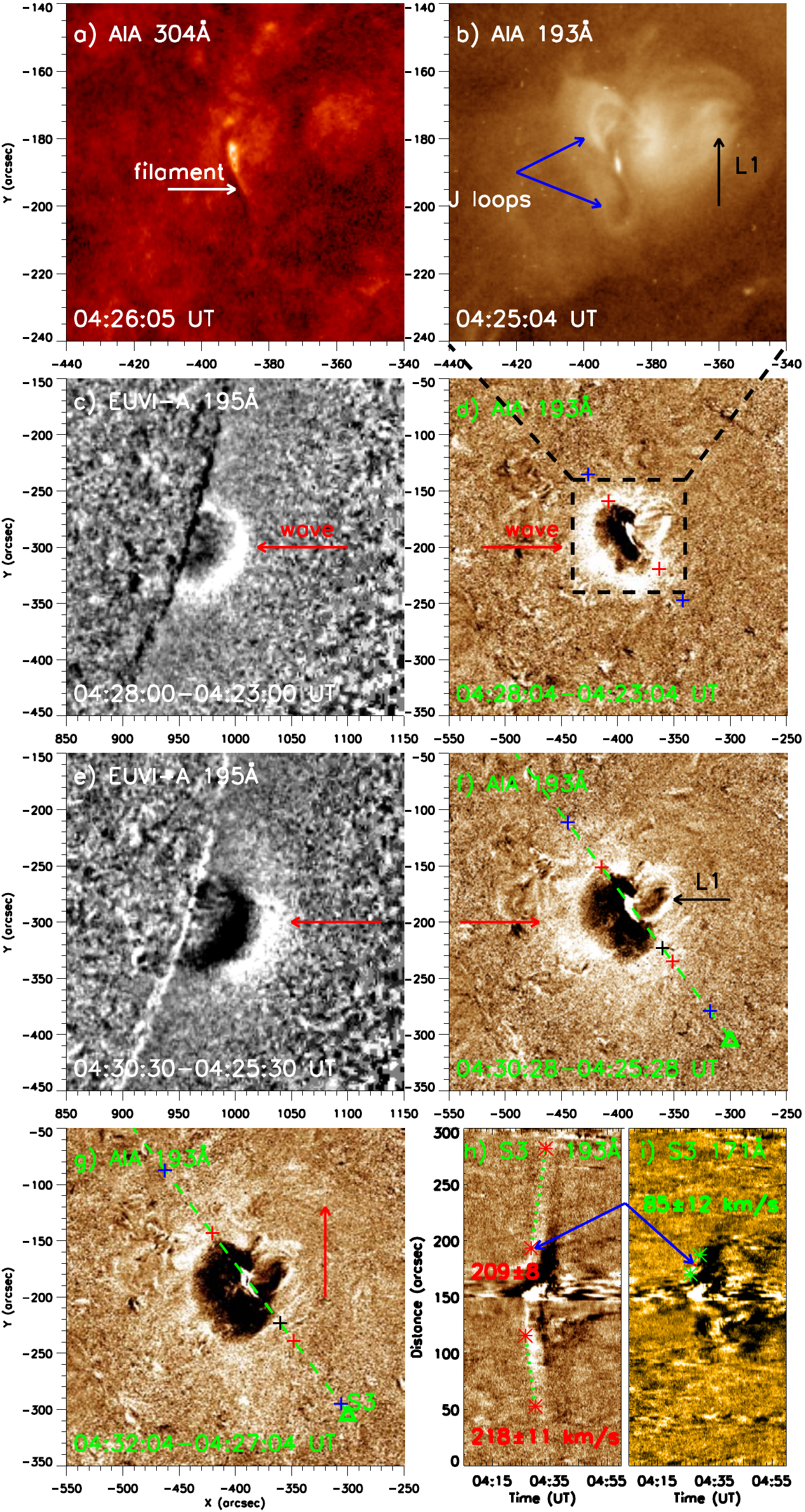}
\caption{The flux-rope-driven wave on 2018 February 24. (a-b) The filament (the white arrow), the J-shaped loops of the sigmoid (blue arrows), and the ambient coronal loops (L1, the black arrow) in direct images of AIA 304 and 193~{\AA}. (c-g) The EUV wave (red arrows) in running-ratio-difference images in EUVI-A 195~{\AA} and AIA 193~{\AA}. The dashed box in (d) indicates the FOV of (a-b). (h-i) The time-slice plots of running-ratio-difference AIA 193 and 171~{\AA} images showing the expanding loops (blue) and the EUV wave (red) along the green dashed line (S3 in (g)). The red and green asterisks connected by dotted lines are used to derive the wave and loop speed.
An animation of panels b, c, and d is available. The animation also includes the evolution of the AIA 171~{\AA} images. In all four panels the video begins at approximately 04:21 UT and ends at 04:39 UT. The video duration is 2 seconds.
\label{f3}}
\end{figure}

\clearpage

\begin{figure}
\epsscale{0.6} \plotone{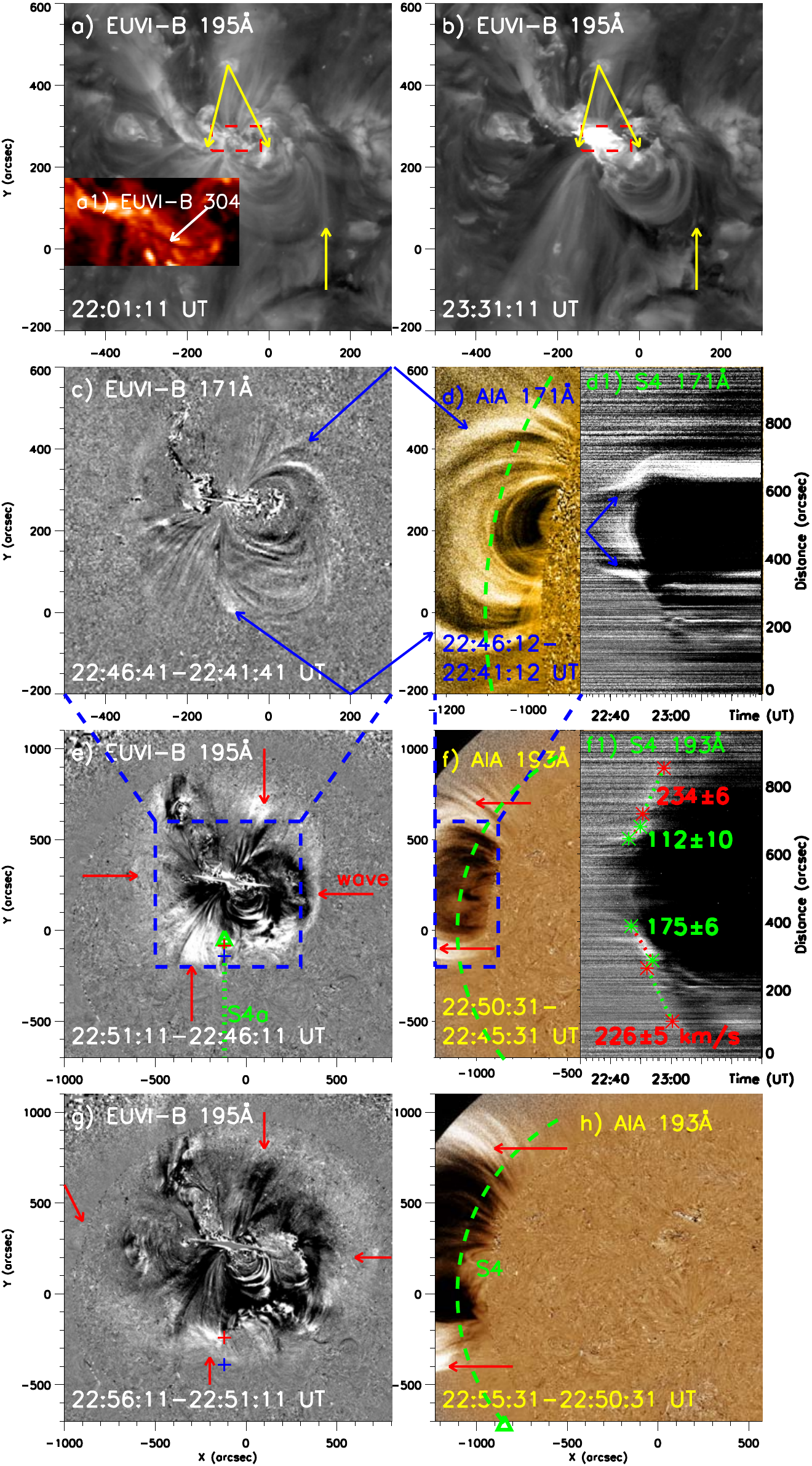}
\caption{The flux-rope-driven wave on 2012 March 26. (a-b) The sigmoid filament (the white arrow in (a1)) and the coronal dimmings (yellow arrows) before and after the sigmoid eruption in direct images of EUVI-B 304 and 195~{\AA}. The dashed boxes indicate the FOV of (a1). (c-h) The expanding loops (blue arrows) and the circular EUV wave (red arrows) in running-ratio-difference images in EUVI-A 171 and 195~{\AA} and AIA 171 and 193~{\AA}. The dashed boxes in (e) and (f) indicate the FOV of (a-c) and the FOV of (d), respectively. (d1 and f1) The time-slice plots of running-ratio-difference AIA 171 and 193~{\AA} images showing the expanding loops (blue arrows) and the EUV wave (red) along the green dashed line (S4 in (h)). The red and green asterisks connected by dotted lines are used to derive the wave and loop speed.
An animation of panels c, d, g, and h is available. In all four panels the video begins at approximately 22:35:30 UT and ends at 22:59:30 UT. The video duration is 2 seconds.
\label{f4}}
\end{figure}

\clearpage

\begin{figure}
\epsscale{1.0} \plotone{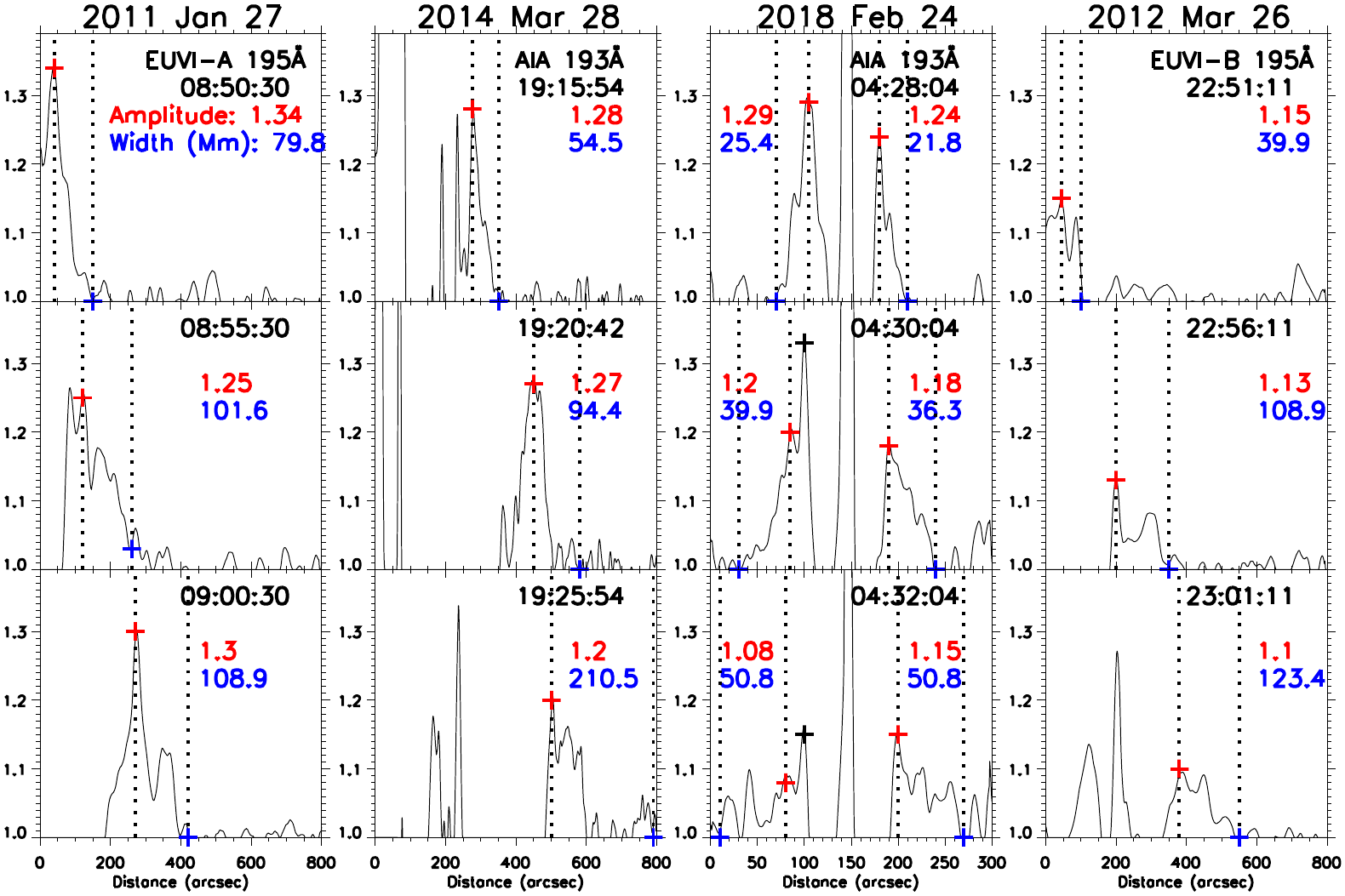}
\caption{The running-ratio intensity profiles of waves in AIA 193~{\AA} or EUVI 195~{\AA} along the selected paths (green dotted lines starting with triangles, S1a, S2, S3, and S4a in Figure 1-4.) The red and blue pluses indicate the peaks and outermost edges of wavefronts, corresponding to the pluses in Figures 1-4, and two neighboring vertical lines show the widths of frontal parts of wavefronts. The red and blue values show the decreasing amplitudes and increasing widths.
\label{f5}}
\end{figure}

\clearpage

\begin{figure}
\epsscale{1.0} \plotone{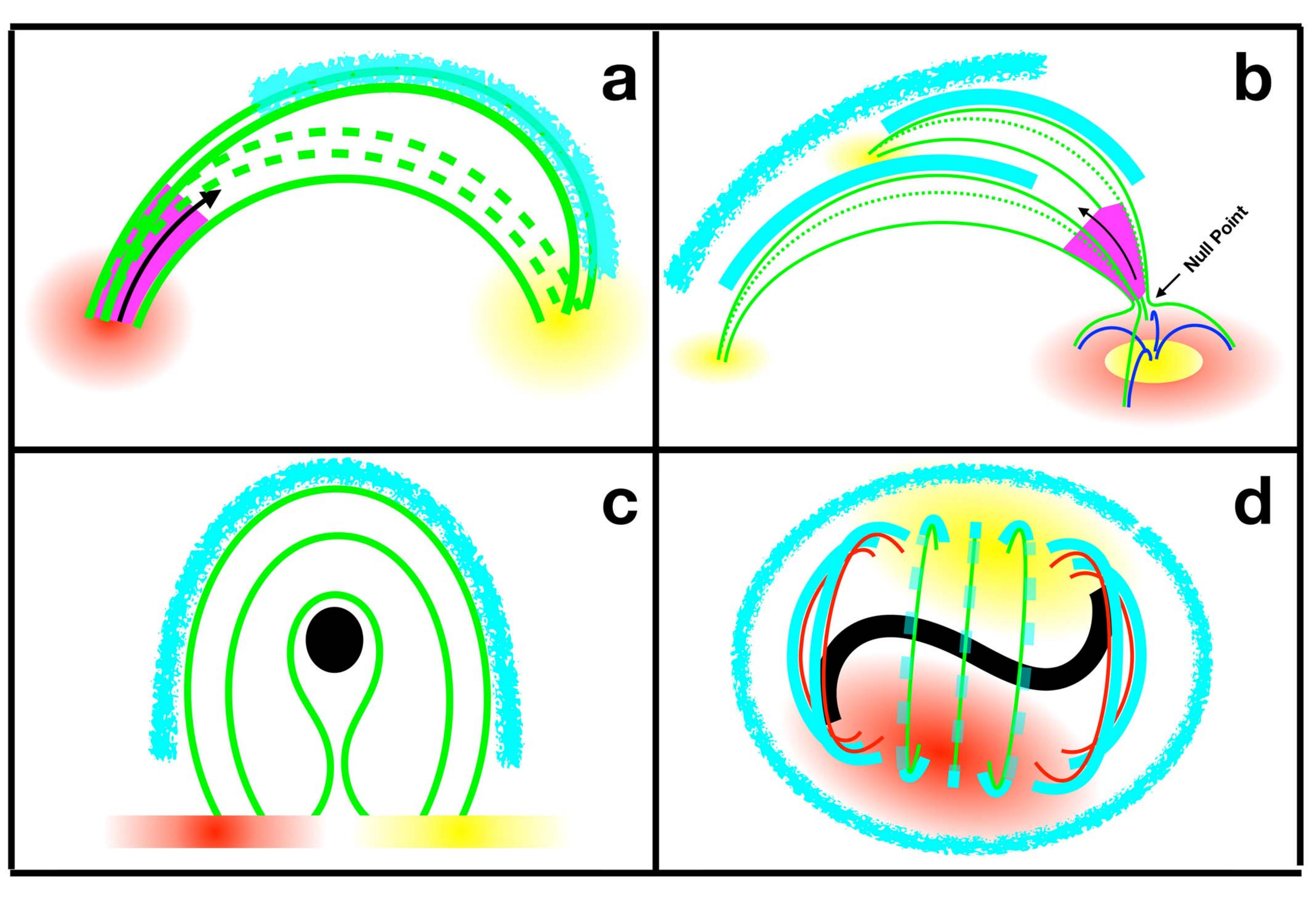}
\caption{The schematic representation showing the scenario of the various wavefront formation for the simple jet (a), the jet in fan-spine-null-point configuration (b), and the flux rope eruption in the side view (c) and top view (d). The jets and flux ropes are indicated by the pink patches and the black circle and S-shaped rope, and their overlying loops are represented by the green and red lines. The dotted or dashed green lines mean the original position of expanding loops. The diffuse blue shades indicate the observed wavefronts of EUV waves, and the smooth blue shades in (b) and (d) show the separated wave components triggered by some isolated loops. The yellow and red shadows indicate the opposite magnetic polarities.
\label{f6}}
\end{figure}


\begin{thebibliography}{}

\bibitem{amari18} Amari, T., Canou, A., Aly, J., Delyon, F. \& Alauzet, F. 2018, Nature, 554, 211

\bibitem{attr07} Attrill, G. D. R., Harra, L. K., van Driel-Gesztelyi, L. \& D{\'e}moulin, P. 2007, \apjl, 656, L101


\bibitem{ballai07} Ballai, I. 2007, \solphys, 246, 177
\bibitem{ballai08} Ballai, I., Douglas, M. \& Marcu, A. 2008, \aap, 448, 1125
\bibitem{ballai05} Ballai, I., Erd{\'e}lyi, R. \& Pint{\'e}r, B. 2005, \apjl, 633, 145L

\bibitem{chen16} Chen, P., Fang, C., Chandra, R. \& Srivastava, A. K. 2016, \solphys, 291, 3195

\bibitem{chen02} Chen, P. F., Wu, S. T., Shibata, K. \& Fang, C. 2002, \apj, 572, 99


\bibitem{chen11} Chen, P. F. \& Wu, Y. 2011, \apjl, 732, 20L

\bibitem{cheng12} Cheng, X., Zhang, J., Olmedo, O., Vourlidas, A., Ding, M. D. \& Liu, Y. 2012, \apjl, 745L, 5

\bibitem{downs12} Downs, C., Roussev, I. I., van der Holst, B., Lugaz, N. \& Sokolov, I. V. 2012, \apj, 750, 134

\bibitem[Howard et al. 2008]{howard08} Howard, R. A., Moses, J. D., Vourlidas, A., et al. 2008, Space Sci. Rev., 136, 67


\bibitem[Kaiser et al. (2008)]{kaiser08} Kaiser, M. L., Kucera, T. A., Davila, J. M., St. Cyr, O. C., Guhathakurta, M., \& Christian, E. 2008, Space Sci. Rev., 136, 5

\bibitem{kwon13} Kwon, R.-Y., Ofman, L., Olmedo, O., Kramar, M., Davila, J. M., Thompson, B. J. \& Cho, K.-S. 2013, \apj, 766, 55

\bibitem{lemen12} Lemen, James R., Title, Alan M., Akin, David J., Boerner, Paul F. et al. 2012, \solphys, 275, 17

\bibitem{li12} Li, T., Zhang, J., Yang, S. \& Liu, W. 2012, \apj, 746, 13

\bibitem{liu10} Liu, W., Nitta, N. V., Schrijver, C. J., Title, A. M. \& Tarbell, T. D. 2010, \apjl, 723, 53L

\bibitem{liu14} Liu, W. \& Ofman, L. 2014, \solphys, 289, 3233

\bibitem{long13} Long, D. M., Williams, D. R., R{\'e}gnier, S. \& Harra, L. K. 2013, \solphys, 288, 567
    
\bibitem{long17} Long, D. M., Bloomfield, D. S., Chen, P. F., et al. 2017, \solphys, 292, 7


\bibitem[Moses et al. (1997)]{mose97} Moses, D., Clette, F., Delaboudini\`ere, J.-P., et al. 1997, \solphys, 175, 571

\bibitem{olme12} Olmedo, O., Vourlidas, A., Zhang, J. \& Cheng, X. 2012, \apj, 756, 143

\bibitem{patsou12} Patsourakos, S. \& Vourlidas, A. 2012, \solphys, 281, 187

\bibitem{pesnell12} Pesnell, W. Dean, Thompson, B. J., Chamberlin, P. C. et al. 2012, \solphys, 275, 3

\bibitem[Scherrer et al. 2012]{sche12} Scherrer, P. H., Schou, J., Bush, R. I., Kosovichev, A. G., et al. 2012, \solphys, 275, 207

\bibitem[Schrijver \& De Rosa (2003)]{schrijver03} Schrijver, C. J., \& De Rosa, Marc L. 2003, \solphys, 212, 165

\bibitem{shen13} Shen, Y., Liu, Y., Su, J., Li, H., Zhao, R., Tian, Z., Ichimoto, K. \& Shibata, K. 2013, \apjl, 773, 133L

\bibitem{shen18b} Shen, Y., Liu, Y., Liu, Y. D., Su, J., Tang, Z. \& Miao, Y. 2018, \apj, 861, 105

\bibitem{shen18a} Shen, Y., Tang, Z., Miao, Y., Su, J. \& Liu, Y. 2018, \apjl, 860, 8

\bibitem{thom98} Thompson, B. J., Plunkett, S. P., Gurman, J. B., Newmark, J. S., St. Cyr, O. C. \& Michels, D. J. 1998, Geophysical Research Letters, 25, 2465

\bibitem[Veronig et al. (2010)]{vero10} Veronig, A. M., Muhr, N., Kienreich, I. W., Temmer, M., \& Vr{\v s}nak, B. 2010, \apjl, 716, 57L

\bibitem[Wang \& Liu (2012)]{wang12} Wang, H., \& Liu, C. 2012, \apj, 760, 101

\bibitem{wang09} Wang, H., Shen, C. \& Lin, J. 2009, \apj, 700, 1716

\bibitem{warm15} Warmuth, A. 2015, Living Reviews in Solar Physics, 12, 3

\bibitem{warm11} Warmuth, A. \& Mann, G. 2011, \aap, 532, A151

\bibitem{warm01} Warmuth, A. Vr{\v s}nak, B., Aurass, H., \& Hanslmeier, A. 2001, \apj, 560, 105L

\bibitem{wills07} Wills-Davey, M. J., DeForest, C. E., \& Stenflo, J. O. 2007, \apj, 664, 556

\bibitem{zheng18} Zheng, R., Chen, Y., Feng, S., Wang, B. \& Song, H. 2018, \apjl, 858, 1L

\bibitem{zheng12a} Zheng, R., Jiang, Y., Yang, J., Bi, Y., Hong, J., Yang, B. \& Yang, D. 2012, \aap, 541, 49

\bibitem{zheng12b} Zheng, R., Jiang, Y., Yang, J., Bi, Y., Hong, J., Yang, B. \& Yang, D. 2012, \apj, 747, 67

\bibitem{zhuk04} Zhukov, A. N. \& Auch\`{e}re, F. 2004, \aap, 427, 705

\bibitem{zong17} Zong, W. \& Dai, Y. 2017, \apjl, 834, 15L

\end{thebibliography}
\end{document}